\documentclass{article} 

\usepackage{blindtext, rotating}
\usepackage{natbib}
\usepackage{multirow}
\usepackage{url,lineno}
\usepackage[all]{xy}

\usepackage{amsfonts}
\usepackage{amssymb}
\usepackage{amsbsy} 
\usepackage{amsmath}
\usepackage{mathrsfs}
\usepackage{dsfont}
\usepackage{bbm}
\usepackage{subfigure}

\usepackage{float}
\usepackage{flushend}

\usepackage{scrextend}

\newcommand{\vect}[1]{\boldsymbol{#1}}

\def\Authors{Marina Martinez-Garcia$^{1,2}$, Marcelo Bertalm\'{i}o$^{3}$, Jes\'{u}s Malo$^{1*}$}

\begin{document}
\onecolumn


\title{In Praise of Artifice Reloaded: \\ Caution with subjective image quality databases}
\author{\Authors}

\maketitle


\vspace{-0.4cm}
\begin{abstract}

Subjective image quality databases are a major source of raw data on how the visual system works in \emph{naturalistic environments}.
These databases describe the sensitivity of many observers to a wide range of distortions
(of different nature and with different suprathreshold intensities) seen on top of a variety of natural images.
They seem like a dream for the vision scientist to check the models in realistic scenarios.

However, while these natural databases are great benchmarks for models developed in some other way (e.g. by using the well-controlled \emph{artificial stimuli} of traditional psychophysics), they should be carefully used when trying to fit vision models.
Given the high dimensionality of the image space, it is very likely that some basic phenomenon (e.g. sensitivity to distortions in certain environments) are under-represented in the database.
Therefore, a model fitted on these large-scale natural databases will not reproduce these under-represented
basic phenomena that could otherwise be easily illustrated with well selected artificial stimuli.

In this work we study a specific example of the above statement.
A wavelet+divisive normalization layer of a sensible cascade of linear+nonlinear layers
fitted to maximize the correlation with subjective opinion of observers on a large image quality database
fails to reproduce basic crossmasking.
Here we outline a solution for this problem using artificial stimuli.
Then, we show that the resulting model is also a competitive solution for the large-scale database.
In line with \citep{Rust05}, our report
(misrepresentation of basic visual phenomena in subjectively-rated natural image databases)
is an additional argument \emph{in praise of artifice}.



\vspace{0.5cm}
{\bf Keywords:} Natural stimuli, Artificial stimuli, Subjective image quality databases, Wavelet + Divisive Normalization, Contrast masking.

\end{abstract}

\newpage

\section{\mbox{Introduction}}

In the age of \emph{big data} one may
think that machine learning applied to representative databases
will automatically lead to accurate models of the problem at hand.
For instance, the problem of modeling the
perceptual difference between images showed up in the discussion of
eventual challenges at the NIPS-11 \emph{Metric Learning} Workshop \citep{NIPS11metric}.
However, despite its interesting implications in visual neuroscience, the subjective metric of the image space
was dismissed as a \emph{trivial} regression problem because there are subjectively-rated image quality
databases that can be used as training set for supervised learning.

Subjective image and video quality databases (such as VQEG, LIVE, TID, CID, CSIQ)\footnote{A non exhaustive list of references and links to subjective quality databases includes \citep{VQEG,LIVE6,TID2008,TID2013,CID,CSIQ}.}
certainly are a major source of raw data on how the visual system works in \emph{naturalistic environments}.
These databases describe the sensitivity of many observers to a wide range of distortions
(of different nature and with different suprathreshold intensities) seen on top of a variety of natural images.
They seem like a dream for the vision scientist to check the models in realistic scenarios.

In fact, following a tradition that links the image quality assessment problem in engineering
with human visual system models \citep{Sakrison77,Watson93,Bovik09,Bovik16},
these subjectively rated image databases have been used to fit models coming from classical psychophysics or physiology \citep{Watson02,Laparra10a,Malo10,Bertalmio17}.
Given the similarity between these biological models \citep{Carandini12} and feed-forward convolutional neural
nets \citep{Goodfellow16}, an interesting analogy is possible.
Fitting the biological models to reproduce the opinion of the observers in the
database is algorithmically equivalent to the learning stage in deep networks.
This deep-learning-like use of the databases is a convenient way to train a physiologically-founded architecture
to reproduce a psychophysical goal \citep{LaparraJOSA17,Martinez17}.
When using these biologically-founded approaches, the parameters found have a straightforward interpretation as for
instance the frequency bandwidth of the system or the extent of the interaction between sensors tuned to different features.

On the other hand, pure machine-learning (data-driven) approaches have also been used to predict subjective
opinion. In this case, after extracting features with reasonable statistical meaning or perceptual inspiration,
generic regression techniques (with no biological grounds) are applied \citep{Moorthy2011,Moorthy2010,Saad2010,Saad2012,Saad2014}.

The problem with the above uses of naturalistic databases (either based on biological models or on data-driven models)
is the conventional concern about training sets in machine learning:
\emph{is the training set a balanced representation of the range of behaviors to be explained?}

If it is not the case, the resulting model may be biased by the dataset and it will have generalization problems.
This overfitting risk has been recognized by the authors of image quality metrics based on generic regression \citep{Saad2012}.
Perceptually meaningful architectures impose certain constraints on the flexibility of the model, as opposed to generic regressors. These constraints could be seen as a sort of \emph{Occam Razor} in favour of lower-dimensional models.
However, even in the biologically meaningful cases, there is a risk that the model
found by fitting the naturalistic database misses well-known texture perception facts.
Accordingly, \citep{Laparra10a,Malo10} used artificial stimuli after the learning stage to check
the Contrast Sensitivity Function and some properties of \emph{visual masking}.

Our message here is that large-scale naturalistic databases should be used carefully when trying to fit vision models.
Given the high dimensionality of the image space, it is very likely that some basic phenomenon
(e.g. the visibility of certain distortion in certain environment) is under-represented in the database.
As a result, the model is not forced to reproduce these under-represented phenomena.

In this work we study a specific example of the generalization risk suggested above.
We show that a wavelet+divisive normalization layer of a sensible cascade of linear+nonlinear layers
fitted to maximize the correlation with subjective opinion of observers on a large image quality database
fails to reproduce basic crossmasking.
Here we point out the problem and we outline a solution using well selected artificial stimuli.
Then, we show that the model corrected to account for these extra artificial tests is also a competitive
explanation for the large-scale naturalistic database.

In line with \citep{Rust05}, our results in this work,
namely pointing out the misrepresentation of basic visual phenomena in subjectively-rated natural image databases and the proposed procedure to fix it, are an additional argument \emph{in praise of artifice}.

The paper is organized as follows: Section \ref{naturalvsartificial} illustrates
the intuition
that can be obtained from proper artificial stimuli
as opposed to the not-so-obvious interpretation of natural stimuli.
Section \ref{failure_section} shows that a sensible psychophysical model tuned to maximize the correlation with subjective opinion in a large-scale naturalistic image quality database (surprisingly?) fails to reproduce
basic properties of visual masking illustrated by simple artificial stimuli.
In Section \ref{solution}, we propose a route to solution:
a change in the formulation of the model allows an intuitive solution of the failures on artificial stimuli
while preserving the success on the large-scale naturalistic database.
Finally, as suggested by the failure-and-solution example considered in this work, in Section \ref{discussion}
we discuss the opportunities and precautions of the use of the subjectively rated databases to fit vision models.

\section{Natural versus artificial stimuli}
\label{naturalvsartificial}

\begin{figure}[!b]
	\centering
    \small
    \setlength{\tabcolsep}{2pt}
    \begin{tabular}{c}
    \hspace{-0.0cm} \includegraphics[width=\textwidth]{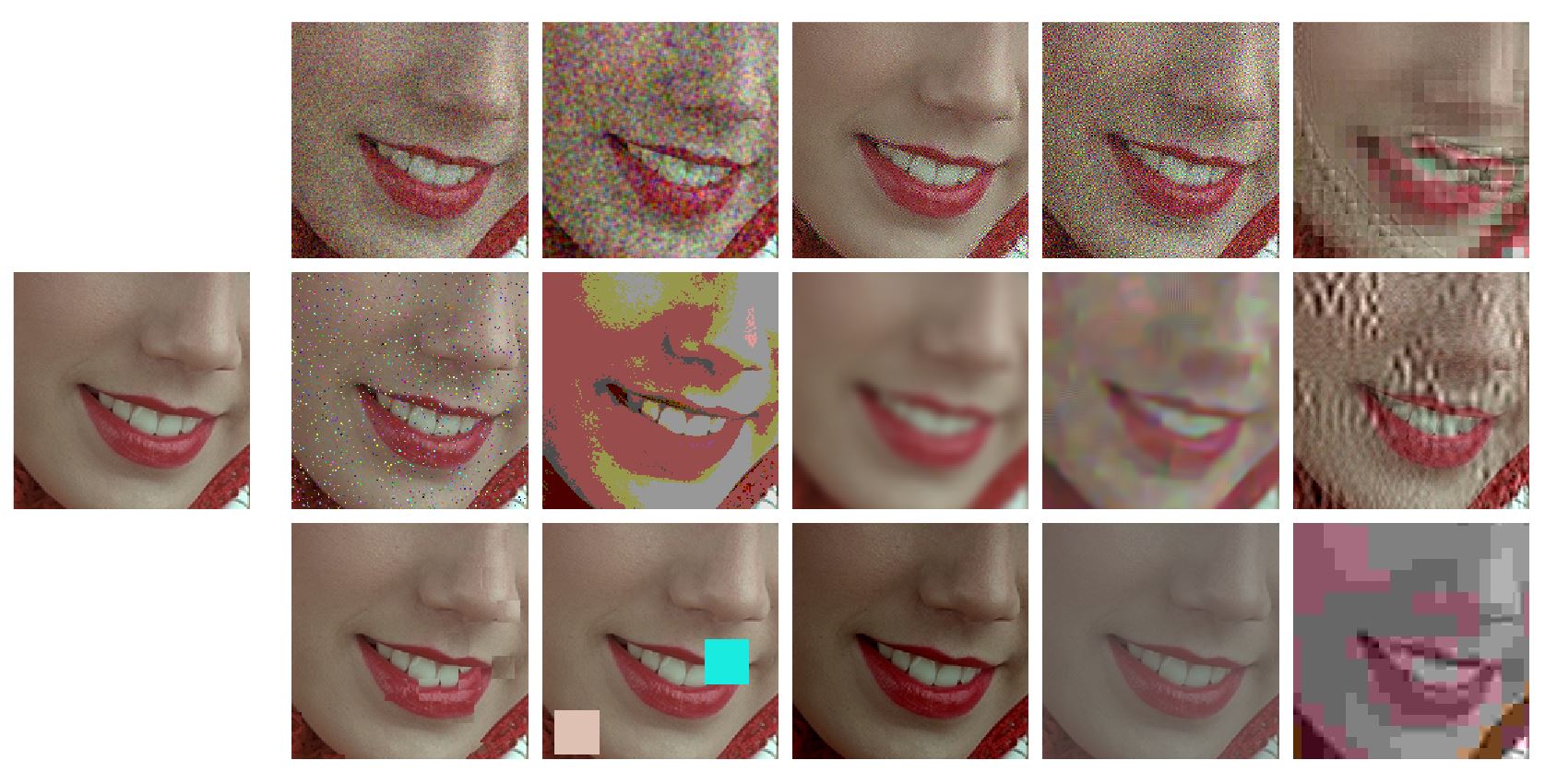} \\
    \end{tabular}
    \vspace{-0.15cm}
	\caption{\emph{Natural scenarios and complex distortions}. Illustrative subset of subjectively rated image quality database: a variety of distortions in naturalistic environment.}\label{naturalstimuli}
    \vspace{-0.15cm}
\end{figure}

Figure \ref{naturalstimuli} shows a representative subset of the kind of patterns subjectively rated in image quality databases.
This specific example comes from the TID2008 database \citep{ponomarenko08}.
In these databases, natural scenes (photographic images with uncontrolled content) are corrupted by noise sources of
different nature (some of them are stationary and signal independent, while others are spatially variant and depend on the background).
Ratings depend on the visibility of the test (the distortion) seen on top of the natural background.
The considered distortions come in different suprathreshold intensities. In some cases these intensities have controlled (linearly spaced)
energy or contrast, but in general, they come from arbitrary scales (e.g. different compression ratio or color quantization coarseness)
with no obvious psychophysical meaning. This is because the motivation of the original databases (e.g. VQEG or LIVE) was the assessment of
distortions occurring in \emph{image processing} applications (e.g. transmission errors in digital communication) and not necessarily to be
a tool for \emph{vision science}. More recent databases include more accurate control of luminance and color of both the backgrounds
and the distortions \citep{CID}, or report the intensities of the distortions in JND units \citep{Chandler14}.
Perceptual ratings in such diverse sets certainly provide a great ground truth to check vision science models in naturalistic conditions.

However, the result of such variety is that the backgrounds and the tests seen on top have no clear interpretation in terms of specific
perceptual mechanisms or controlled statistics in a representation with physiological meaning.
Even though not specifically directed against subjectively rated databases, this was also the main drawback pointed out in \citep{Rust05} against the use of generic natural images in vision science experiments.

In this work we go a step further in that criticism: due to the uncontrolled nature of the natural scenes and the somewhat arbitrary distortions found in these databases, the different aspects of a specific perceptual phenomenon are not fully represented in the database.
Therefore, these databases should be used carefully when training models because this misrepresentation will have consequences when fitting the models.

For instance, let's consider pattern masking \citep{Foley94,Watson97}. It is true that some distortions in the databases introduce relatively more noise in high contrast regions, which seems appropriate to illustrate masking. This is the case of the JPEG or JPEG2000 artifacts, or the so called \emph{masked noise} in the TID database (e.g. the third example in the first row of Fig. \ref{naturalstimuli}).
These deviations on top of high contrast regions are less visible than equivalent deviations of the same energy on top of flat backgrounds.
This difference in visibility is due to the inhibitory effect of surround in \emph{masking} \citep{Foley94,Watson97}. Actually, perceptual improvements of image coding standards critically depend on using better masking models that allow using less bits in those regions \citep{malo2000role,Malo01a,Taubman01,Malo06a}.
Appropriate prediction of the visibility of these distortions in the database should come from an accurate model of texture masking.
However, a systematic set of examples illustrating the different aspects of masking (e.g. crossmasking between different frequencies in different backgrounds) is certainly not present in the databases. Therefore, this phenomenon is under-represented in the database.

\begin{figure}[!b]
	\centering
    \small
    \setlength{\tabcolsep}{2pt}
    \begin{tabular}{c}
    Low-frequency vertical test\\
    \hspace{-0.0cm} \includegraphics[width=\textwidth]{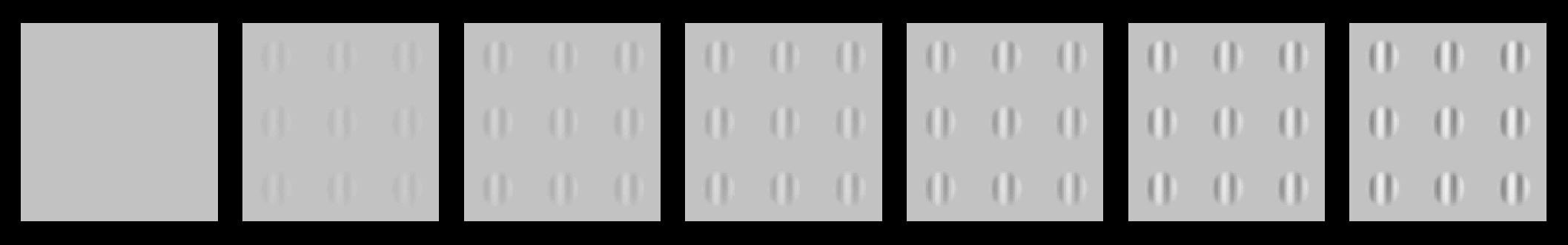} \\[-0.3cm]
    \hspace{-0.0cm} \includegraphics[width=\textwidth]{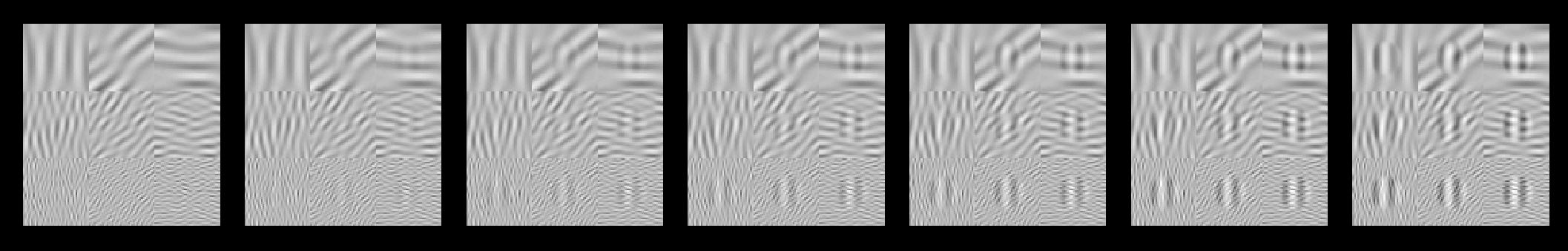} \\
    High-frequency horizontal test\\
    \hspace{-0.0cm} \includegraphics[width=\textwidth]{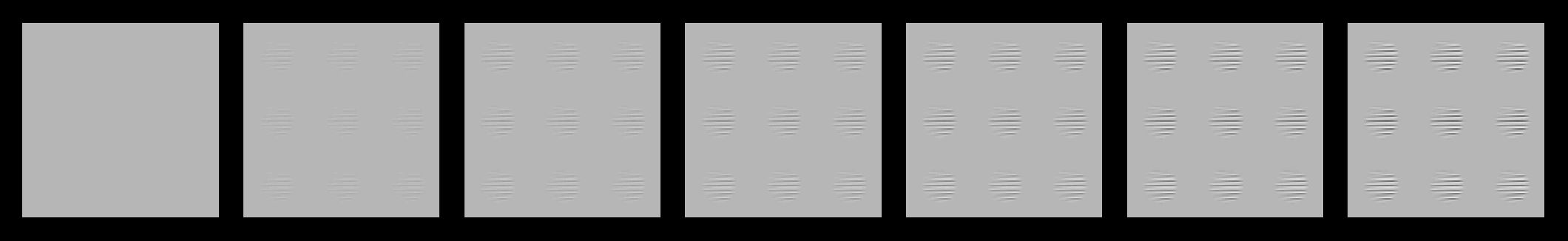} \\[-0.3cm]
    \hspace{-0.0cm} \includegraphics[width=\textwidth]{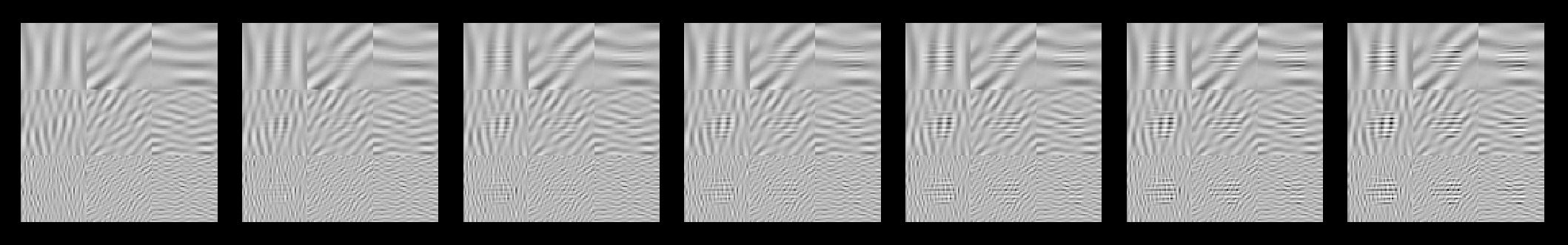}\\
    \end{tabular}
    \vspace{-0.15cm}
	\caption{\emph{Artificial stimuli}.
    Several texture perception phenomena illustrated in a single figure. Frequency sensitivity, nonlinearity and crossmasking (see text).
}\label{artifstimuli}
    \vspace{-0.15cm}
\end{figure}

Fortunately, such basic texture perception facts can be easily illustrated using artificial stimuli.
Artificial stimuli can be designed with a specific perceptual phenomenon in mind, and using patterns
which have specific consequences in models (e.g. stimulation of certain sensors of the model).
Model/phenomenon-based stimuli is the standard way in classical psychophysics and physiology.
Figure \ref{artifstimuli} is an example of the power of well controlled artificial stimuli:
it represents a number of major texture perception phenomena in a single figure.

This figure shows two basic tests (low-frequency vertical and high-frequency horizontal) of increasing contrast from left to right.
These series of tests are respectively shown on top of (a) no background, and (b) on top of backgrounds of controlled frequency and
orientation.
\begin{figure}[!t]
	\centering
    \small
    \setlength{\tabcolsep}{2pt}
    \begin{tabular}{c}
    \hspace{-0.0cm} \includegraphics[width=0.9\textwidth]{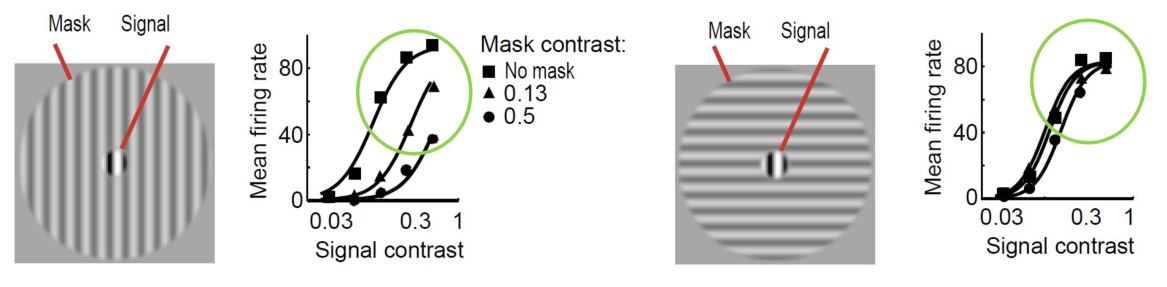} \\
    \end{tabular}
    \vspace{-0.15cm}
	\caption{Experimental response of V1 neurons (mean firing rate) in masking situations.
 Adapted from \cite{Schwartz2001,Cavanaugh2000}. It is important to stress the decay in the response
 when test and mask do have the same spatio-frequency characteristics, as opposed to the case where they do not
 (difference in the circles in green).}\label{psico_effect}
    \vspace{-0.15cm}
\end{figure}

First, of course we can see that the visibility of the tests (or the response of the mechanisms that
mediate visibility) increases with contrast (from left to right).
This is why even the trivial Euclidean distance is positively correlated with subjective opinion of distortion.

Second, the visibility (or the responses) depend(s) on the frequency of the test.
Note that the lower frequency test is more visible than the high frequency test (at reading distance).
This illustrates the effect of the Contrast Sensitivity Function \citep{Campbell68}.

Third, the visibility (or response) increase is non-linear with contrast.
Note that for lower contrasts (e.g. from the second picture to the third in the series) the increase
in visibility is bigger than for higher contrasts (e.g. between the pictures at the right-end).
This means that the slope of the mechanisms mediating the response is high for lower amplitudes and saturates afterwards.
This sort of Weber-like behavior for contrast is a distinct feature of contrast masking \citep{Legge81}.

Finally, the visibility (or response) decreases with the background energy depending on the spatio-frequency
similarity between test and background.
Note for instance that the low frequency test is less visible on top of the low frequency background than
on top of the high frequency background.
Equivalently (important for the example considered throughout this paper) note that the visibility of the
high frequency test behaves \emph{the other way around}: it is bigger on top of the low frequency test.
Moreover (you can easily imagine even though it is not shown in the picture), this \emph{masking} effect
is bigger for bigger contrasts of the background.
This adaptivity of the nonlinearity is a distinct feature of the \emph{masking} effect \citep{Foley94,Watson97},
and more importantly, it is a distinct feature of real neurons \citep{Carandini94,Carandini12} with regard to the
simplified neurons used in deep learning \citep{Goodfellow16}.

As a result, just by looking at this figure, one may imagine how the visibility (or response) curves versus the contrast of the test
should be for the series of stimuli presented. Figure \ref{psico_effect} shows an experimental example of the kind of response curves versus contrast
obtained in actual neurons in masking situations.
Even these qualitative ideas may be used to discard models that do not reproduce
the expected behavior (that do not agree with what we are seeing).

\begin{figure}[!t]
	\centering
    \small
    \setlength{\tabcolsep}{2pt}
    \begin{tabular}{c}
    \hspace{-0.0cm} \includegraphics[width=\textwidth]{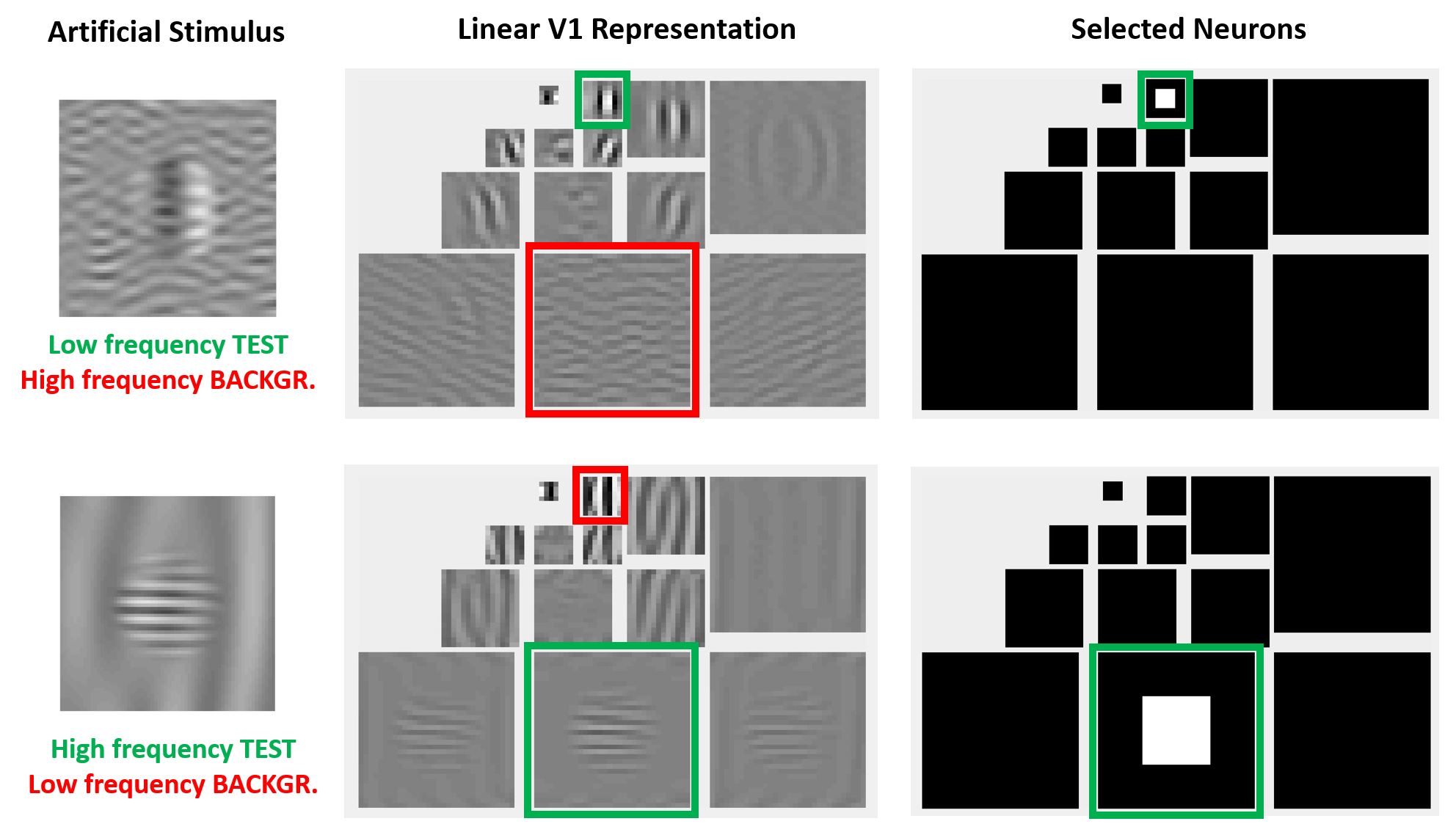} \\
    \end{tabular}
    \vspace{-0.15cm}
	\caption{\emph{Advantages of artificial stimuli}.
    Model-related construction of stimuli simplify the reproduction of results form model outputs and straightforward interpretation of results.
}\label{wavstimuli}
    \vspace{-0.15cm}
\end{figure}

More importantly, the relative visibility of these artificial stimuli can also be used to
intuitively tune the parameters of a model to better reproduce the expected behavior (the visible behavior).
This can be done because these artificial stimuli were crafted to have a clear interpretation in
a standard model of texture vision: a set of V1-like wavelet neurons
(oriented filters tuned to different frequency scales) interacting through cross-inhibitions.
Figure \ref{wavstimuli} illustrates this fact: note how the test patterns considered in the figure mainly stimulate a specific subband of a 4-scale 4-orientation
steerable wavelet pyramid \citep{Simoncelli92}, which is a commonly used formalism to model V1 sensors.
As a result, it is easy to select the set of sensors that will drive the visibility descriptor in the model
(highlighted wavelet coefficients in the diagrams at the right in Fig. \ref{wavstimuli}).

The same intuitive energy distribution over the pyramid is true for the backgrounds,
which stimulate the corresponding subband (scale and orientation).
As a result, given the distribution of test and backgrounds in the pyramid, one may imagine
cross-inhibition strategies to lead to the required decays in the response.

These intuitions obtained from artificial model-oriented stimuli (intuitive response curves and eventual-crossmasking
strategies between subbands) are fundamental both to criticise the results obtained from \emph{blind learning from a database},
and to propose intuitive improvements of the model.

\section{The problem: unexpected failure of model after optimization}
\label{failure_section}

In vision science, cascades of \emph{Linear}+\emph{Nonlinear} transforms are very successful
in modeling a number of perceptual experiences~\citep{Carandini12}.
For example perceptions of color, motion and spatial texture are tightly related
to L+NL models of similar functional form~\citep{Brainard05,Simoncelli98,Watson97}.

Specifically, \citep{Carandini12} suggest a standard program for a spatial vision model based on a cascade
of isomorphic L+NL layers addressing in turn (1)~brightness, (2)~contrast computation, (3)~frequency sensitivity and energy masking,
and (4)~multi-scale oriented analysis and frequency masking.

In \citep{Martinez17} we presented the mathematics required to get the parameters of such canonical cascade according
to two different strategies:
(a) the Jacobian of the transform w.r.t. the image allows novel psychophysics
based on building model-related artificial stimuli to select the right parameters of the architecture, e.g.
using \emph{Maximum Differentiation} techniques \citep{Wang08,Malo15}; and
(b) the Jacobian of the transform w.r.t the parameters allows parameter learning according to a specific goal.
The latter is formally similar to the training phase of deep networks \citep{Goodfellow16},
with a fundamental difference: in our case the proposed goal was related to a sensible psychophysical
task (predicting the visibility of distortions in natural environments) and we were training a physiologically plausible
architecture: the canonical Divisive Normalization.

Thanks to these two strategies (based on \emph{artificial} and \emph{naturalistic} stimuli) the program suggested
in \citep{Carandini12} became a computational model where these four stages were psychophysically tuned to work
together for the first time. The detailed formulation of this base-line canonical model
(let us call it \textbf{Model~A}) is given in the Methods Section \ref{modelA}.

In \citep{Martinez17} some layers of \textbf{Model~A}, namely the contrast and energy masking (layers 2nd and 3rd), were determined using the model-based artificial-stimuli strategy; while other layers, namely the brightness and the wavelet + divisive normalization (layers 1st and 4th), were determined using the natural-database strategy.
In fact, since that work was focused on the mathematics, the use of both strategies was just an arbitrary illustration of the
correctness of the proposed analytical results.

Beyond the formal correctness of the results, which were extensively checked in different ways in \citep{Martinez17},
here we analyze the practical effects and the eventual limitations of the strategy based on learning in natural uncontrolled
large-scale databases. To do that, in this work we focus only on the 4th layer: wavelet + divisive normalization.

\subsection{Success of \emph{"Model~A"} in naturalistic databases}

The layer of \textbf{Model~A} intended to account for scale and orientation analysis and frequency masking is pretty standard:
it follows a canonical L+NL structure \citep{Carandini12}, where the linear stage is a 4-scale, 4-orientation steerable wavelet transform \citep{Simoncelli92} of the output of the previous layer; and the nonlinear stage is a divisive normalization transform  \citep{Watson97} of the wavelet coefficients. Assuming that the output of the wavelet filter-bank is the vector $\vect{y}$, and assuming that the vector of energies of the coefficients is obtained by coefficient-wise rectification and exponentiation, $\vect{e} = |\vect{y}|^\gamma$, the vector of responses after divisive normalization in the last layer of \textbf{Model~A} is:
\begin{equation}
      \vect{x} = sign(\vect{y}) \odot \frac{\vect{e}}{\vect{b} + H \cdot \vect{e}}
      \label{DN_A}
\end{equation}

\noindent where $\odot$ stands for element-wise Hadamard product and the division is also an element-wise Hadamard quotient where the energy of each linear response is divided by a linear combination of the energies of the neighboring coefficients in the wavelet pyramid. This linear combination (that attenuates the response) is given by the matrix-on-vector product $H \cdot \vect{e}$.
Note that, for simplicity, in Eq.~\ref{DN_A} we omitted the indices referring to the 4th layer (as opposed to the more verbose formulation in the Methods Section \ref{modelA}).

The $i$-th row of this matrix, $H$, tells us how the responses of neighbor sensors in the vector $\vect{e}$ attenuate the response of the $i$-th sensor in the numerator, $e_i$. The attenuating effect of these linear combinations is moderated by the semisaturation constants in vector $\vect{b}$.

\begin{figure}[!t]
	\centering
    \small
    \setlength{\tabcolsep}{2pt}
    \begin{tabular}{c}
    \hspace{-0.0cm} \includegraphics[width=\textwidth]{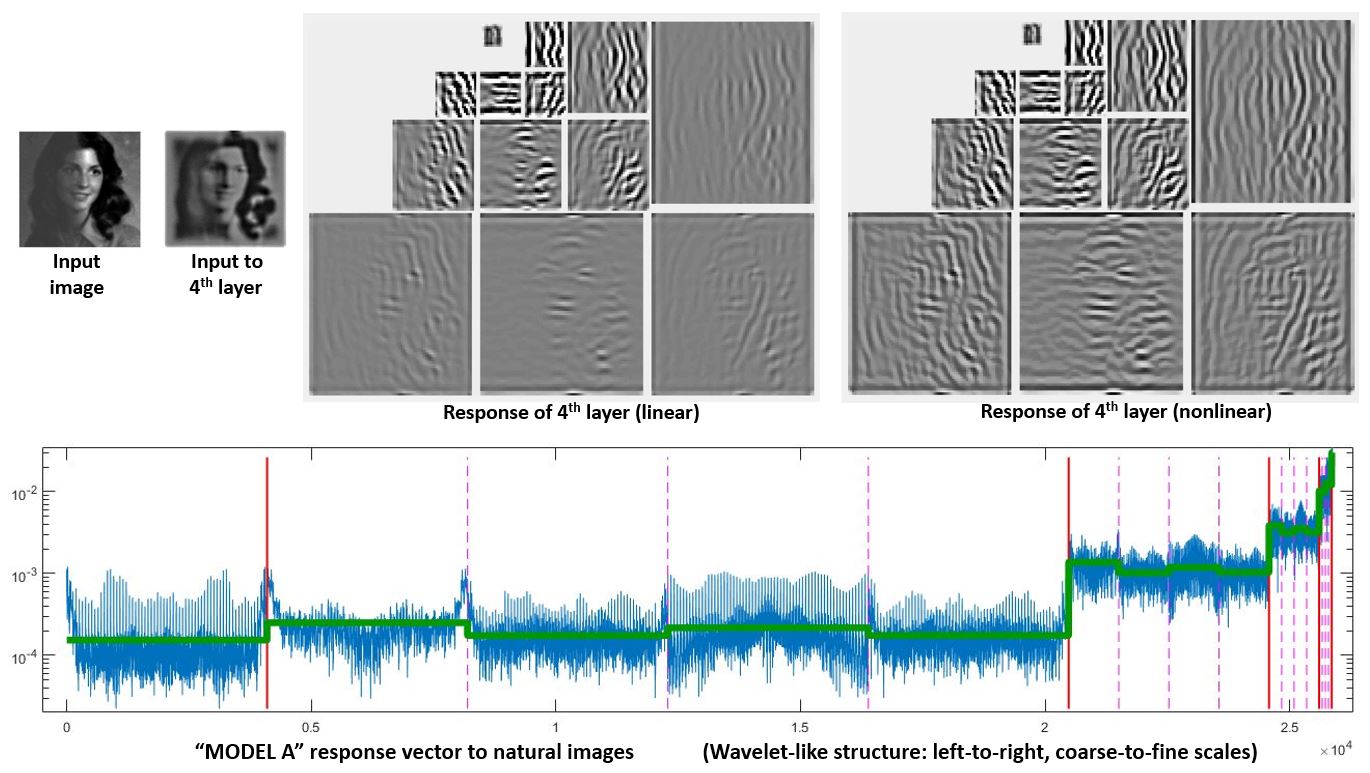} \\
    \end{tabular}
    \vspace{-0.15cm}
	\caption{\emph{Response of Model~A to natural images}.
    Given a luminance distribution (input image), the initial layers of the model (retina-to-LGN) compute a filtered version of brightness contrast with adaptation to lower contrasts due to divisive normalization (input to 4th layer). Finally, the linear part of the 4th layer computes a multi-scale / multi-orientation decomposition and then, these responses nonlinearly interact as given by Eq. \ref{DN_A}. The structure of a representative vector of responses depicted at the bottom is relevant to understand the assumed interactions and the eventual modifications that may be required.
    As usual in the wavelet literature \citep{Simoncelli90}, data in the vector are organized from high-frequency (fine scales at the left) to low-frequency (coarse scales at the right). Vertical lines in red indicate the limits of the different scales. Within each scale, the dashed lines in pink indicate the limits of the different orientations. The different coefficients within each scale/orientation block correspond to different spatial locations.
    The average line in green shows the average amplitude per subband for a set of natural images.
    As discussed in the text, this specific energy distribution per scale is relevant for the good performance of the model.
}\label{nat_wav_stimuli}
    \vspace{-0.15cm}
\end{figure}

The structure of these vectors and matrices is relevant to understand the behavior on the stimuli.
First, one must consider that all the vectors, $\vect{y}$,  $\vect{e}$, and  $\vect{x}$, have wavelet-like
structure. Fig. \ref{wavstimuli} shows this subband structure for specific artificial stimuli and Fig. \ref{nat_wav_stimuli} shows it for natural stimuli.

It is important to stress that the specific distribution of responses of natural images over the
subbands of the response vector (green line in Fig.\ref{nat_wav_stimuli}) is critical to reproduce the
good behavior of the model on the database.
Note that this is not a regular (linear) wavelet transform, but the (nonlinear) response vector.
Therefore, this distribution tells us \emph{both} about the statistics of natural images and about the behavior of the visual system.
On the one hand, it is true that natural images have relatively more energy in the low-frequency end.
But, on the other hand, it is visually relevant that the response of sensors tuned to the high frequency details
keep certain relation to (is much lower than) the response of the sensors tuned to the low frequency details.
The latter is in line with the different visibility of the artificial stimuli of different frequency shown in
Fig. \ref{artifstimuli}, and it is probably due to the effect of the CSF filter in earlier stages of the model.
This is important because keeping this relative magnitude between subbands is crucial to have good alignment with
subjective opinion in the large-scale database.

\begin{figure}[!t]
	\centering
    \small
    \setlength{\tabcolsep}{2pt}
    \begin{tabular}{c}
    \hspace{-0.0cm} \includegraphics[width=\textwidth]{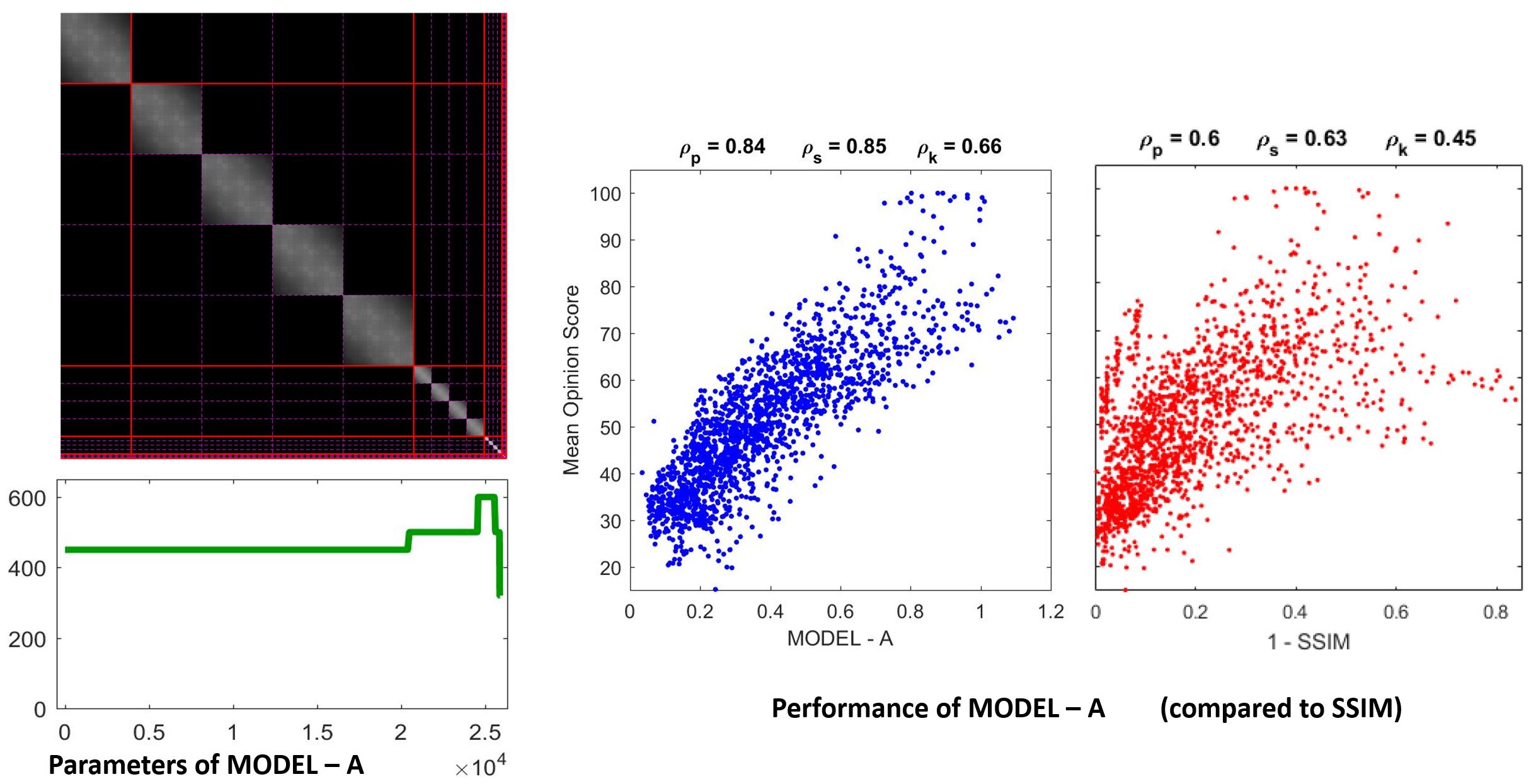} \\
    \end{tabular}
    \vspace{-0.15cm}
	\caption{\emph{Parameters of MODEL-A (left) and performance on large scale naturalistic database (right)}.
    The parameters are: the interaction kernel $H$ (matrix on top), and the semisaturation per subband vector, $\vect{b}$.
}\label{kernels}
    \vspace{-0.15cm}
\end{figure}

Once we understand the spatio-frequency structure of the response vector\footnote{The $i$-th coefficient has a 4-dimensional spatio-frequency meaning, $i \equiv ( \vect{p}_i, f_i, \phi_i)$, where $\vect{p}$ is a two-dimensional location, $f$ is the modulus of the spatial frequency, and $\phi$ is orientation.}, we may take a look at the optimal
interaction kernel and semisaturation to maximize the correlation with mean opinion score over the large-scale
naturalistic database.

In \citep{Martinez17} we assumed Gaussian interaction kernels following \citep{Watson97}.
We only assumed intraband interactions
to keep equations simple, while still illustrating the generality of the analytical results.
Moreover, simplicity was not the only reason: there are reports \citep{Laparra10a,Malo10} that suggest
that the intraband relations are the most relevant interactions
in terms of improving the correlation with subjective opinion.
Optimization of the width and amplitude of the Gaussian kernel in each subband as well as
the semisaturation parameter in each subband\footnote{This is referred to as \emph{optimization phase I} in \citep{Martinez17}.
Even though in \emph{optimization phase II} using the full variability in $\vect{b}$ correlations were higher,
we restrict ourselves to \emph{optimization phase I} because we want to keep the number of parameters small.
In phase I optimization only 1/25 of the database was used in the training.}
led to the results referred to as \textbf{Model~A} in Fig. \ref{kernels}.
Spatial-only intraband relations lead to symmetric block diagonal kernels.
Optimization acted on the width and amplitude of these kernels per subband.
Similarly, optimization lead to bigger semisaturation for low frequencies (except for the low-pass residual).

The performance of the resulting model on the naturalistic database is certainly good: compare
the correlation of \textbf{Model~A} with subjective opinion in Fig. \ref{kernels} as opposed to the widely
used Structural SIMilarity index \citep{wang04}, in red, included for useful reference.
Given the difference in correlation with regard to SSIM, one can certainly say that \textbf{Model~A} is
\emph{highly successful} in predicting the visibility of uncontrolled distortions seen on
naturalistic backgrounds.

\subsection{Relative failure of \emph{"Model~A"} with artificial stimuli}

Despite the reasonable formulation of this base-line \textbf{Model~A} and its successful performance in reproducing
subjective opinion in large-scale naturalistic databases, a simple simulation with the kind of artificial stimuli
presented in Section \ref{naturalvsartificial} shows that it does not reproduce all the aspects
of basic visual masking.

Specifically, we computed the response curves of the highlighted neurons in Fig. \ref{wavstimuli}
for low-frequency and high-frequency tests like those illustrated in Fig. \ref{artifstimuli}
as a function of their contrast. We considered four different contrasts for the background.
Different orientations of the background (vertical, diagonal and horizontal) were also considered.

\begin{figure}[!t]
	\centering
    \small
    \setlength{\tabcolsep}{2pt}
    \begin{tabular}{c}
    \hspace{-0.0cm} \includegraphics[width=0.9\textwidth]{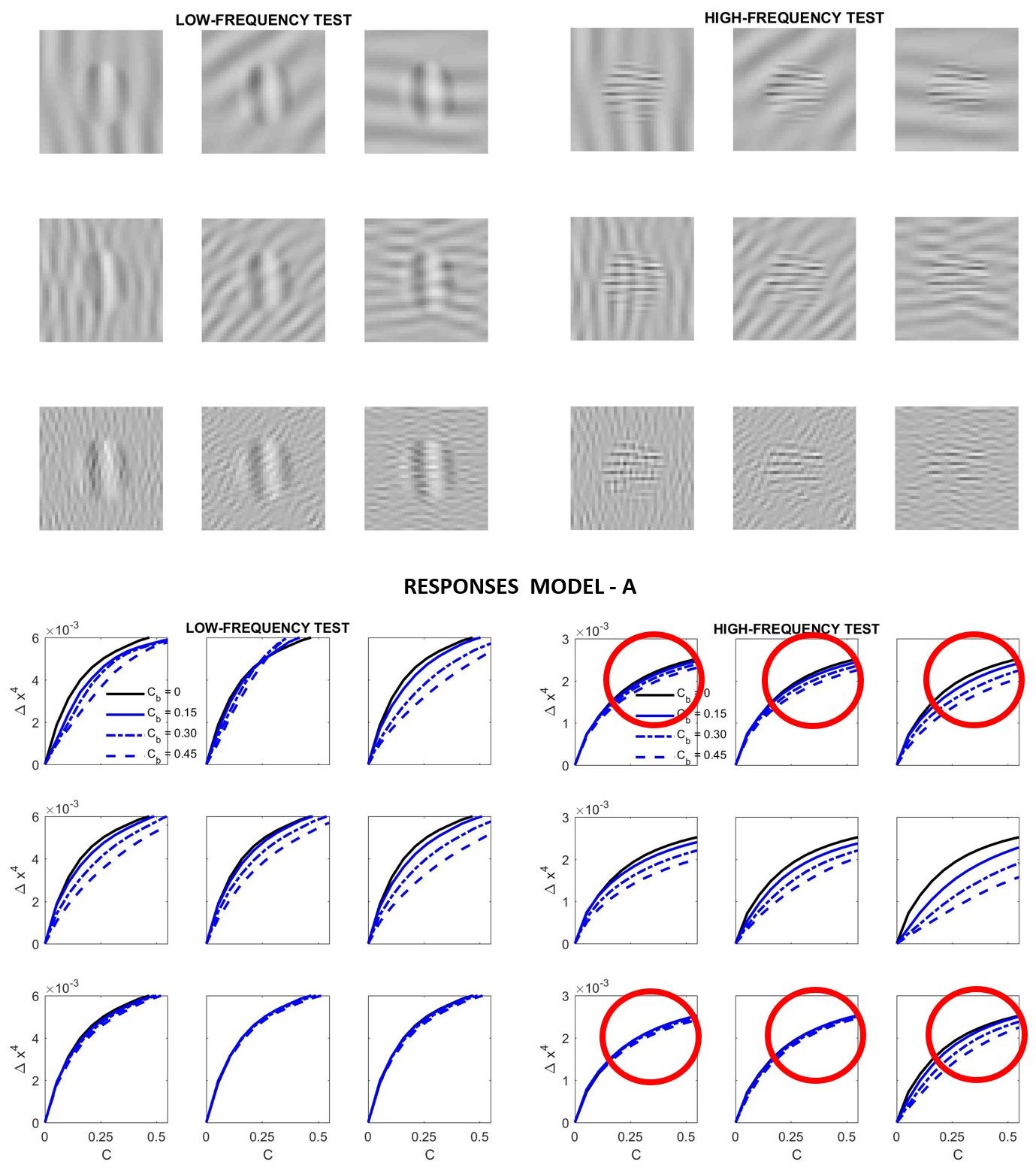} \\
    \end{tabular}
    \vspace{-0.15cm}
	\caption{\emph{Relative success and failures of optimized model}.
    Model-related construction of stimuli simplify the reproduction of results form model outputs and straightforward interpretation of results.
}\label{failure}
    \vspace{-0.15cm}
\end{figure}

Figure \ref{failure} presents the results of such simulation.
This figure highlights some of the good features of \textbf{Model~A}, but also its shortcomings.

On the positive side we have the following.
First, (as expected) the response increases with contrast.
Second, (as expected) the response for the low frequency test is bigger than the response for the
high frequency test.
Third, (as expected) the response saturates with contrast.
And also, increasing the contrast of the background decreases the response.

However, \emph{contrarily to what we can see when looking at the artificial stimuli},
the response for the high frequency test \emph{does not} decay more on top of high frequency backgrounds.
While the decay behavior is roughly OK for the low-frequency test, definitely it is not OK for the high-frequency test.
Compare the decays of the signal at the circles highlighted in red in Fig. \ref{failure}:
the response of the sensors tuned to high-frequency test decays by the same amount when they are presented
on top of low-frequency backgrounds than when the background also has high-frequency.
The model is failing here.

\section{A route to solution: fine-tune for artificial while preserving performance for natural}
\label{solution}

As stated above, the benefits of artificial model-oriented stimuli are not limited to the identification
of failures.
When models are interpretable (as is \textbf{Model~A} as opposed to pure data-driven models),
model-based stimuli also suggest intuitive ways to modify the model in case of shortcomings.

In this section we heuristically modify the model provided by blind optimization (\textbf{Model~A}) by applying
the intuition obtained from the artificial stimuli and the associated simulation of response
curves. We try to fine-tune for the artificial stimuli while preserving the good performance for the natural database.
%
%
%

\subsection{First guess: generalize kernel and reduce semisaturation}

The failure observed in Fig. \ref{failure} consists of having a too strong influence of low-frequency activity into the response of high-frequency sensors.
Therefore, the first natural idea is including extra inter-band relations
in the only-intra-band kernel, $H$, of \textbf{Model~A}.
Following \citep{Watson97} we may consider an interaction mediated by a Gaussian function which depends on the distance between the location of the sensors, $H_{\vect{p}}$,
and is also modulated by two additional Gaussian interactions depending on the difference of scales and orientations:  $H_{f}$ and $H_{\phi}$.
As in \textbf{Model~A}, the normalization of each row of the kernel (the normalization of each interaction
neighborhood) is controlled by a diagonal matrix $\mathds{D}_{\vect{c}}$, which contains the vector of normalization constants, $\vect{c}$,
in the diagonal.
Considering that symmetry in Gaussians may not be perceptually appropriate,
here we introduce two extra matrices ($C_{\textrm{int}}$ and $\mathds{D}_{\vect{w}}$) to break this symmetry in the generalized kernel $H_G$, in case a more detailed control is required:
\begin{equation}
      H_G = \mathds{D}_{\vect{c}} \cdot \left[ H_{\vect{p}} \odot H_{f} \odot H_{\phi} \odot C_{\textrm{int}} \right] \cdot \mathds{D}_{\vect{w}}
      \label{new_kernel_eq}
\end{equation}
where $C_{\textrm{int}}$ is a subband-wise full matrix, and $\mathds{D}_{\vect{w}}$ is a diagonal matrix with vector $\vect{w}$ in the diagonal.
Note that $\mathds{D}_{\vect{w}}$ applies column-wise weights on the final kernel, or equivalently, it selectively
weights the energy of the subbands in the input vector $\vect{e}$. This means that it can be used to moderate the effect of the
low-frequency (which is too big in \textbf{Model~A}).
More importantly, one could act on a specific block of the full matrix, $C_{\textrm{int}}$, if the relation between two specific
subbands should be modified.

Moreover, there is a second relevant intuition: modifications in the kernel may be ineffective if the semisaturation constants are too high.
Note that the denominator of Divisive Normalization, Eq. \ref{DN_A}, is a balance between the linear combination $H \cdot \vect{e}$
and the vector $\vect{b}$. This means that some elements of $\vect{b}$ should be reduced for the subbands where we want to act.
Increasing the corresponding elements of vector $\vect{c}$, leads to a similar effect.

With these intuitions one can start playing with $H$ and $\vect{b}$.
However, while the effect of the low-frequency is easy to reduce using the above ideas (thus solving the problem highlighted in red in Fig.~\ref{failure}), the relative amplitude between the responses to low and high frequency inputs is also easily lost.
This quickly ruins the CSF-like behavior and reduces the performance on the large-scale database.
We should not lose the relative amplitudes of the responses of \textbf{Model~A} to natural images (i.e. green lines in Fig. \ref{nat_wav_stimuli}) to keep its good performance.
Unfortunately \textbf{Model~A} is unstable under this kind of modifications making it difficult to tune.

\subsection{New formulation: general and easier-to-tune model}

In order to solve the above scaling problems,
it is convenient to include a global scaling factor that sets the dynamic range of the responses,
particularly the relative amplitude of the different scales. See the new term in brackets:
\begin{equation}
      \vect{x} = sign(\vect{y}) \odot \left[  \vect{\kappa} \odot \frac{ \vect{b} + H_G \cdot \vect{e}^\star}{\vect{e}^\star} \right] \odot \frac{\vect{e}}{\vect{b} + H_G \cdot \vect{e}}
      \label{DN_B}
\end{equation}
In this modified expression, which we will refer to as \textbf{Model~B}, the response still follows a nonlinear divisive normalization,
but when the input $\vect{e}$ arrives to the \emph{reference} value $\vect{e}^\star$, we ensure that the scale of the output is given by
the vector $\vect{\kappa}$.
The reference value, $\vect{e}^\star$, should be related to the dynamic range of the \emph{input} signal to this neural layer.
The global scaling factor, $\vect{\kappa}$, should be related to the dynamic range we want to impose in the \emph{output} of this neural layer.


Accordingly, $\vect{e}^\star$ is something we can compute from the previous layers of the model (that we are not modifying here).
For instance, $\vect{e}^\star$ may be an average amplitude (per subband) for natural images in stage $\vect{y}^4$, in the notation of Section \ref{modelA},
or it can be signal adaptive (as the average \emph{anchor} luminance in the first layer of the model).
See the details for a convenient computation of these averages in Section \ref{modelB}.

In the same vein, we propose to set $\vect{\kappa}$ (the dynamic range of the output) as the average per subband of the final response of \textbf{Model~A} for natural images (i.e. the green curve in Fig. \ref{nat_wav_stimuli}).

The detailed formulation of this layer of \textbf{Model~B}, namely matrix formulation of the forward transform, Jacobian and inverse,
is given in Section \ref{modelB} for the interested reader. The code with a Matlab implementation of both \textbf{Model~A} and \textbf{Model~B}
(that also includes numerical check of the analytical results presented in Section \ref{modelB}) is available at {\tiny{\verb"http://isp.uv.es/docs/BioMultiLayer_L_NL_a_and_b.zip"}}.

With these modifications, \textbf{Model~B} in Eq. \ref{DN_B} with the kernel of Eq. \ref{new_kernel_eq},
is both general and has a stabilized and sensible dynamic range at the output.
Then, \textbf{Model~B} is ready for the intuitive modifications considered above.


\subsection{New results}

\begin{figure}[!t]
	\centering
    \small
    \setlength{\tabcolsep}{2pt}
    \begin{tabular}{c}
    \hspace{-1cm} \includegraphics[width=1.1\textwidth]{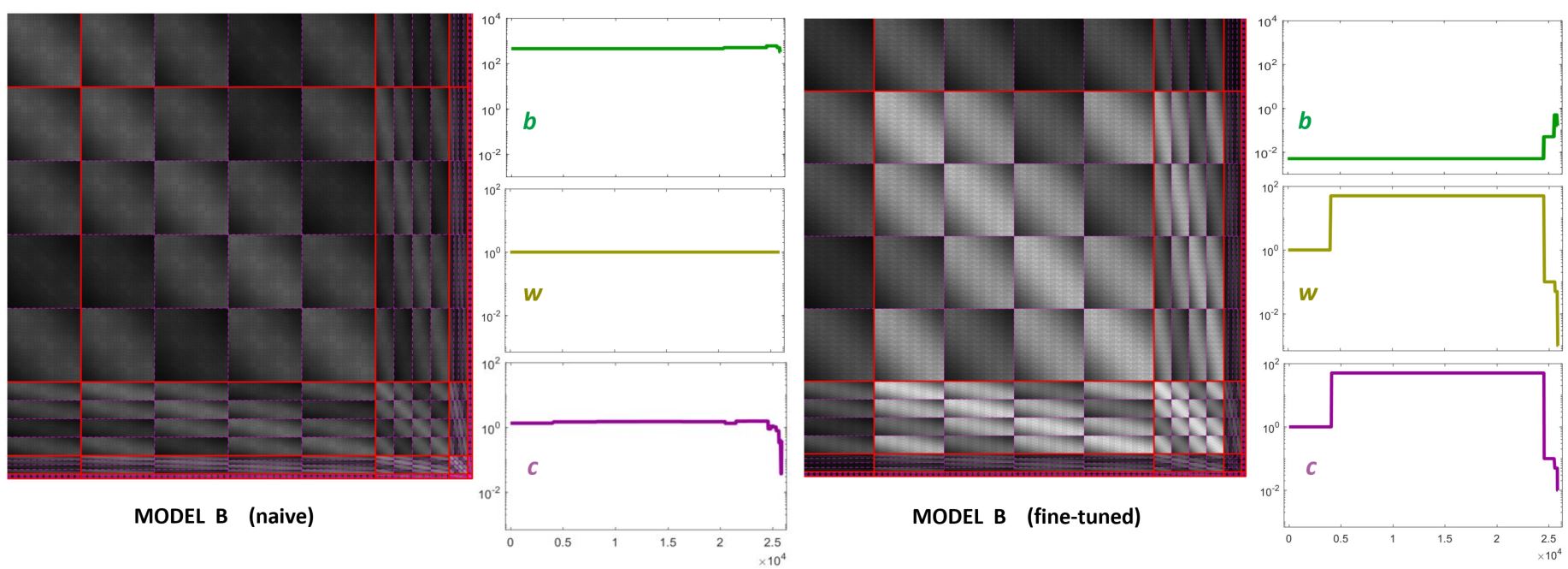} \\
    \end{tabular}
    \vspace{-0.15cm}
	\caption{\emph{Parameters of the modified models}.
    Left panel shows the interaction matrix and the semisaturation vector of the first guess for Model - B. It is called \emph{naive} because the semisaturation and amplitudes of the kernel are imported from the optimized case. The panel at the right shows the corresponding parameters for the fine-tuned version of Model B.
}\label{new_parameters}
    \vspace{-0.15cm}
\end{figure}

The starting point of our heuristic exploration is a straightforward translation of \textbf{Model~A} into \textbf{Model~B}.
We will refer to this as \textbf{Model~B naive}.
This starting point consists of importing the values of the parameters from \textbf{Model~A} except for the modulations depending on the scale
and orientation. We assumed reasonable interaction lengths of one octave (for scales) and 30 degrees (for orientation).
We used no extra weights to break the symmetry ($C_{\textrm{int}} = \mathds{1}$ is an all-ones matrix, and $C_{\vect{w}} = \mathbbm{1}$ is the identity).
And the values for $\vect{c}$ and $\vect{b}$ also come from \textbf{Model~A}.
The parameters of \textbf{Model~B naive} are shown in Fig. \ref{new_parameters} (left panels).
The idea of this starting point, \textbf{Model~B naive}, is reproducing the behavior of \textbf{Model~A} to
build on from there.

\begin{figure}[!t]
	\centering
    \small
    \setlength{\tabcolsep}{2pt}
    \begin{tabular}{c}
    \hspace{-0.0cm} \includegraphics[width=0.88\textwidth]{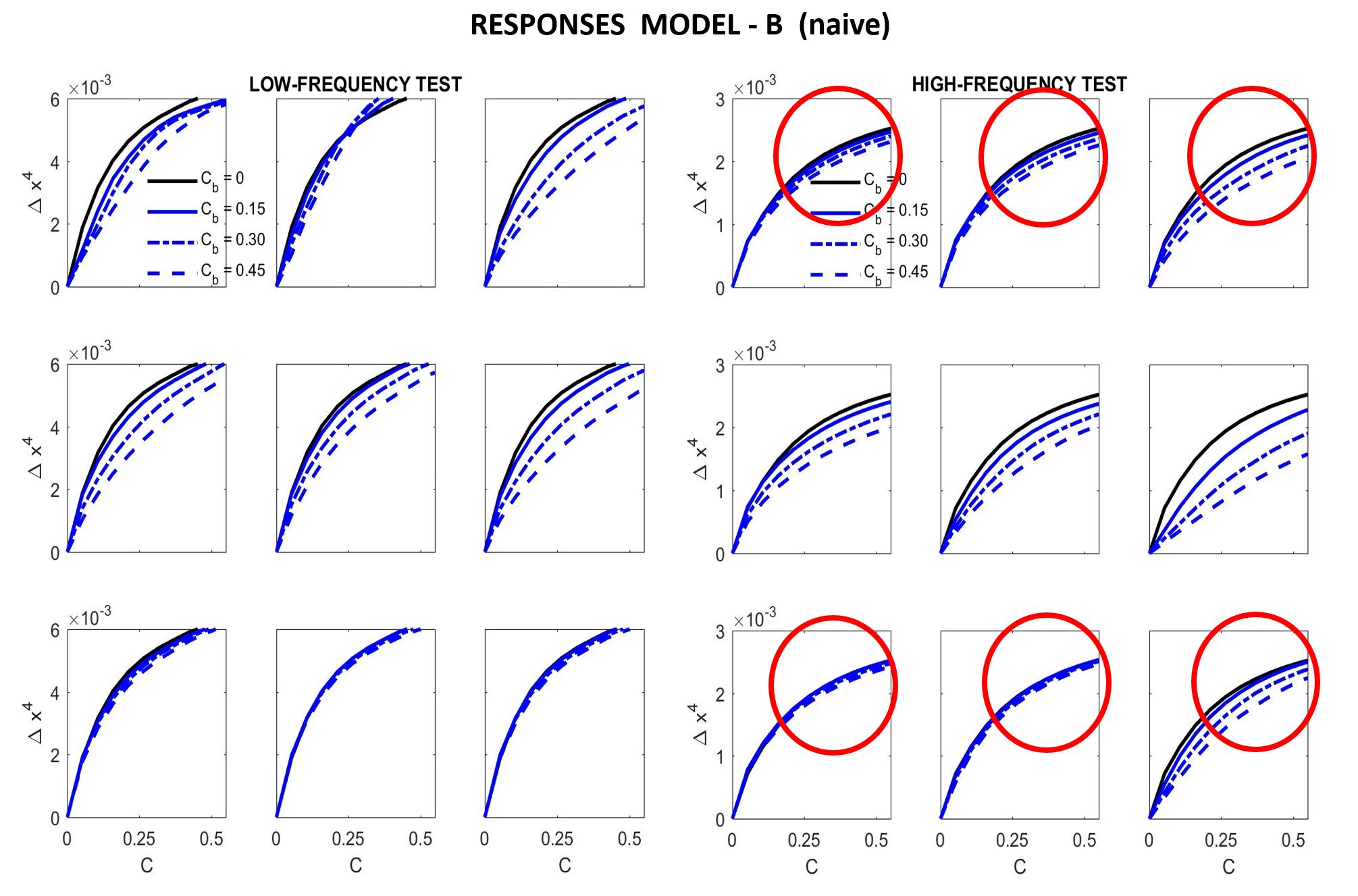} \\
    \includegraphics[width=0.88\textwidth]{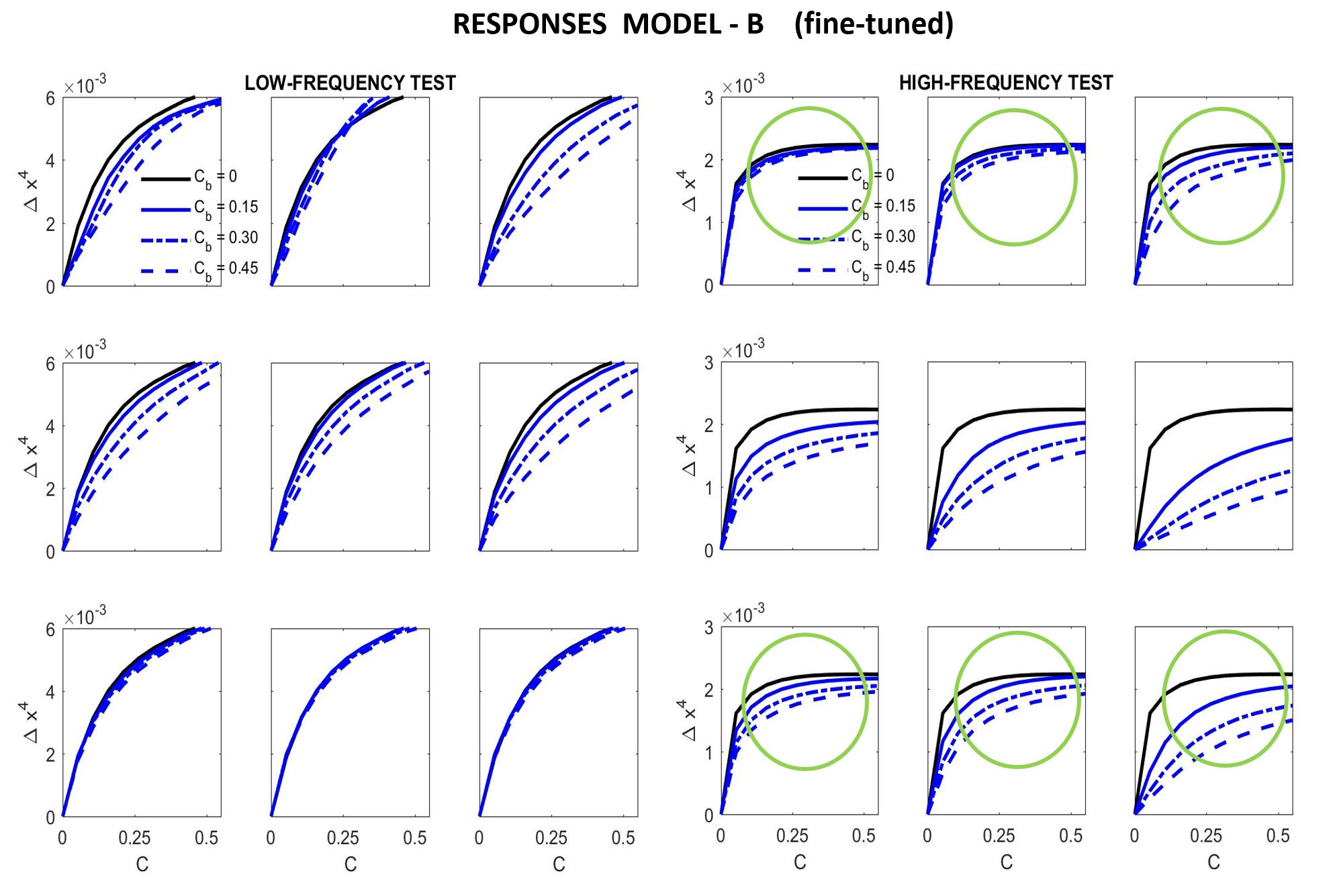} \\
    \end{tabular}
    \vspace{-0.15cm}
	\caption{\emph{Responses of the modified models for the artificial stimuli}.
}\label{successA}
    \vspace{-0.15cm}
\end{figure}

\begin{figure}[!t]
	\centering
    \small
    \setlength{\tabcolsep}{2pt}
    \begin{tabular}{cc}
    \hspace{-0.0cm} \includegraphics[width=0.35\textwidth]{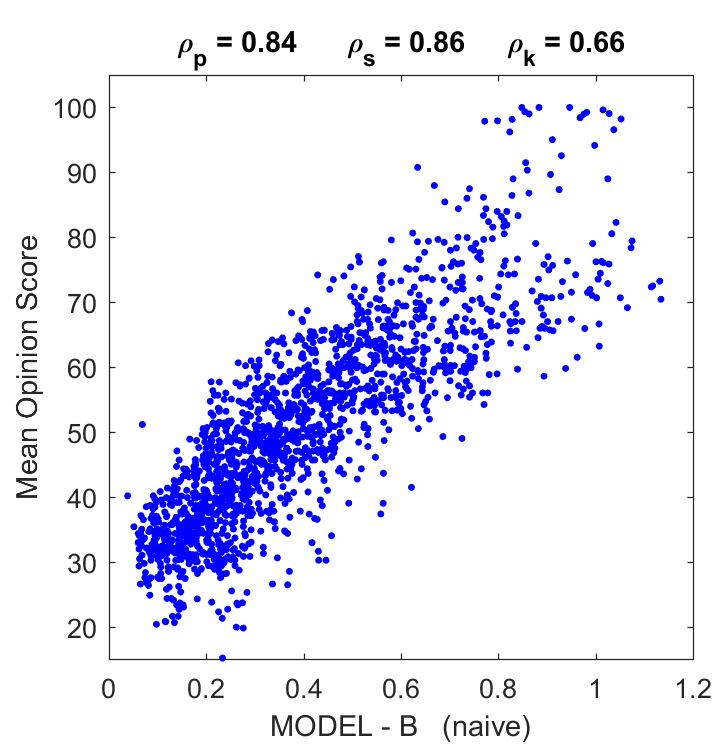} \hspace{0.5cm}&\hspace{0.5cm}
    \hspace{-0.0cm} \includegraphics[width=0.35\textwidth]{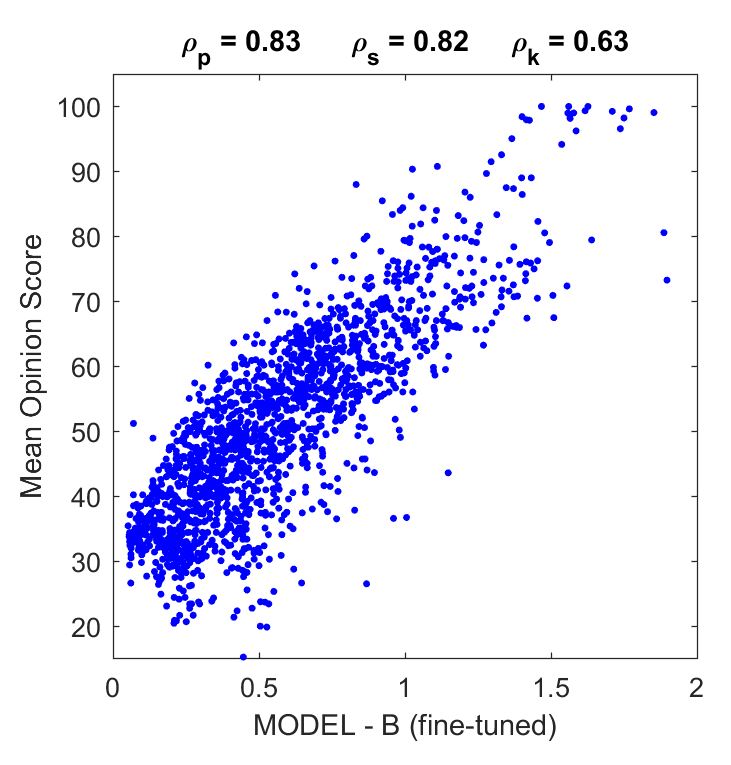} \end{tabular}
 \caption{\emph{Performance of modified models on the natural image database}.
}\label{successN}
    \vspace{-0.15cm}
\end{figure}

Results in Figures \ref{successA} and \ref{successN} show that \textbf{Model~B naive} certainly reproduces the
behavior of \textbf{Model~A} (both the success in the natural image database and the relative failure with artificial stimuli).

Figure \ref{new_parameters} (right panel) shows the fine-tuned parameters according to the heuristic described above.
Note that we strongly reduced $\vect{b}$ and we applied bigger reductions for the high-frequency bands (which corresponds to the sensors we want to fix).
In the same vein we increased the values for the global scale of the kernels of high frequencies $\vect{c}$ while reducing substantially these amplitudes
for low-frequencies to preserve previous behavior (which was roughly OK).
Finally, and more importantly, we moderated the effect of the low-frequencies in masking by using small weights for the low-frequency scales in $\vect{w}$
and while increasing the values for high frequency. Note how this reduces the columns corresponding to the low-frequency subbands in the final kernel $H_G$, and the other way around for the high-frequency scales. This implies a bigger effect of high-frequency backgrounds in the attenuation of high-frequency sensors and reducing the effect of the low-frequency.

Results in Fig. \ref{successA} show that this fine-tuning qualitatively fixes the problem detected in \textbf{Model~A}, which was also present in
\textbf{Model~B naive}. We successfully modified the response of high-frequency sensors (see the decay in the green circles compared to the behavior in the red circles). And we introduced no major difference in the low-frequency responses (which already were roughly correct).

Moreover, Figure~\ref{successN} shows that the fine-tuned version of \textbf{Model~B} not only works better for artificial stimuli
but it also preserves the success in the natural image database.
This is probably due to the positive effect of fixing the relative magnitude of the responses as in \textbf{Model~A}.

It is interesting to stress that this fix didn't require any extra weight in $C_{\textrm{int}}$, which remained an all-ones matrix.
We only operated row-wise and column-wise with the diagonal matrices $\mathds{D}_{\vect{c}}$ and $\mathds{D}_{\vect{w}}$ respectively.

%
%
%

\section{Discussion}
\label{discussion}

The relevant question is: \emph{is the failure of \textbf{Model~A} something that we could have expected?}
And the unfortunate answer is, \emph{yes}: the failure is not surprising given the (almost necessarily) unbalanced nature of large-scale databases.
Note that it is not only that \textbf{Model~A} is somewhat
rigid\footnote{It is true that \textbf{Model~A} only included intra-band relations, but note also that, even though we wanted to
introduce more general kernels in \textbf{Model~B} for future developments,
the solution to the problem considered here basically came from including $\mathds{D}_{\vect{w}}$ in $H$ (not from the specific cross-subband weights).
The other ingredients, $\vect{b}$ and $\vect{c}$ were already present in \textbf{Model~A}.},
the fundamental problem is that despite the specific phenomenon is present in the database,
it is not in the right amount (with enough frequency and intensity) to force the model to
reproduce it in the \emph{learning} stage.

Of course, this problem is hard to solve because it is not obvious to decide in
advance the kind of phenomena (and the right amount of each one) that should be present in the database(s):
as a result, databases are almost necessarily unbalanced and biased by the original intention of the
creators of the database.

Here we made a full analysis (problem and route-to-solution) on texture masking,
but note that focus on masking was just one important but arbitrary example to stress
the main message.
There are equivalent limitations affecting other parts of the optimized model that may come from
the specific features of the database.
For instance, the luminance-to-brightness transform (first layer in models A and B) is
known to be strongly nonlinear and highly adaptive \citep{Stiles82,Fairchild13}.
It can be modeled using the canonical divisive normalization \citep{Abrams07,Brainard05}
but also other alternative nonlinearities \citep{Cyriac16}, and this nonlinearity has been shown to
have relevant statistical effects \citep{Laughlin83,Laparra12,Laparra15,KaneBertalmio16}.
However, when fitting layers 1st and 4th simultaneously to reproduce subjective opinion over the naturalistic
database in \citep{Martinez17}, even though we found a consistent increase in correlation, in the end,
the behavior for the first layer turned out to be almost linear. The constant controlling the effect of the anchor luminance
($\beta$ in the kernel of Eq. \ref{layer1} in Section \ref{modelA}) turned out to be very high.
Again, one of the reasons for this result may be that the low dynamic range of the database did not
require a stronger nonlinearity at the front-end (given the rest of the layers).

%
%
%
%

Incidentally, even though this is a deeper philosophical issue still under debate \citep{Castelvecchi16,Bohannon17},
the case studied here is not only a praise of artificial stimuli, but also a praise of \emph{interpretable models}.
When models are interpretable, it is easier to fix their problems from their failures on synthetic model-related stimuli.
For example, the problem and solution we described here is not limited to \emph{divisive} models of adaptation.
Following \citep{Bertalmio17}, it may be also applied to other interpretable models such as the \emph{subtractive} Wilson-Cowan equations \citep{Wilson72,BertalmioCowan09},
by tuning the parameters that describe the relations between sensors.
This would have been more difficult, if possible at all, with data-driven methods.

\subsection{Using naturalistic databases is always a problem?}

Our criticism of naturalistic databases because their eventual unbalance and the problem in interpreting complicated stimuli
in terms of models does not mean that we claim for an absolute rejection of these naturalistic databases. Not at all.
The case we studied here only suggests that one should not use the databases \emph{blindly} as the only source of information,
but in appropriate combination with well-selected artificial stimuli. 

The use of carefully selected artificial stimuli may be considered as a safety-check of biological plausibility.
Of course, our intention with the case studied here was not exhausting the search possibilities to claim that we obtained
some sort of optimal solution. Instead, we just wanted to stress the fact that using the appropriate stimuli
it is easy to propose modifications of the model that go in the right (biologically meaningful) direction,
and still represent a competitive solution for the naturalistic database.
This is an intuitive way to jump to other local minima which may be more biologically plausible
in a very different region of the parameter space.

In fact, a sensible procedure would be alternating different learning epochs using natural and artificial
data: while the large-scale naturalistic databases coming from the \emph{image processing} community
may enforce the main trends of the system, the specific small-scale artificial stimuli coming from the \emph{vision science}
community will fine-tune that first order approximation so that the resulting model has the appropriate
features revealed by more specific experiments.
In this context, standardization efforts such as those done by the CIE and the OSA organizations
(e.g. the data supporting the standard color observer \citep{CIE31,Stockman17} and the
standard spatial observer \citep{Modelfest}) are really important to make this double-check.

From a more general perspective, we feel that \emph{image processing} applications
do have a fundamental interest in \emph{visual neuroscience} because these applications
put into a broader context the relative relevance of the different phenomena described by classical psychophysics or physiology.
For instance, one can check the variations in performance by testing vision models of different
complexity (e.g. with or without this or that nonlinearity accounting for some specific perceptual effect/ability).
This approach oriented to check different perceptual modules in specific applications has been applied in
image quality databases \citep{Watson02}, but also in other domains such as perceptual image and video
compression \citep{malo2000role,Malo06a,malo2000importance,Malo01a}, or in perceptual image denoising
and enhancement \citep{Gutierrez06,Bertalmio14}.
These different applications show the relative relevance of improvements in masking models with regard to better CSFs or including more sensible motion estimation models in front of better texture perception models.

\subsection{Are all the databases created equal?}

The case analyzed in this work illustrates the effect of (naively) using a database where texture masking is
probably under-represented. The lesson to learn is that one has to take into account the phenomena
for which database was created (or, equivalently, the absence of specific phenomena to address).

With this in mind, one could imagine what kind of artificial stimuli are needed to improve the results.
Or alternatively, which other naturalistic databases (more focused on other kind of perceptual behavior) are required as
complementary check.

Some examples to illustrate this point: databases with controlled observation distance or accurate chromatic
calibration such as \citep{CID} are more appropriate to set the spatial frequency bandwidth of the models in achromatic and
chromatic channels. Databases with spectrally controlled illumination pairs \citep{Laparra12,Gutmann14,Laparra15} are
appropriate to address chromatic adaptation models. Databases with high-dynamic range \citep{Parraga16,Ebrahimi15}
will be more appropriate to point out the need of the nonlinearity of brightness perception.
Finally, databases where visibility of incremental patterns was carefully controlled in contrast terms \citep{Chandler14}
are the best option to fit masking models as opposed to generic subjectively-rated image distortion databases.

\subsection{Final remarks}

Previous literature \citep{Rust05} criticised the use of too complex natural stimuli
in vision science experiments because the statistics of such stimuli are difficult
to control and conclusions may be biased by the interaction between this poorly controlled input
and the complexities of the neural model under consideration.

In line with such precautions on the use of natural stimuli, here we make a different point:
the general criticism to blind use of machine learning in large-scale databases
(related to the proper balance in the data) also applies when using subjectively rated image databases to fit vision models.
Using a variety of natural scenarios and distortions cannot guarantee that specific behaviors
are properly represented, thus remaining hidden in the vast amount of data.
In such situation, models that seem to have the right structure may miss these basic phenomena.
Instead, it is easier to fix these problems with simpler model-oriented artificial stimuli in
qualitative but illustrative experiments.

The case study considered here suggests that artificial stimuli (motivated by specific phenomena or
by features of the model) may help both to (a) stress the problems that remain in models fitted to
unbalanced natural image databases, and (b) to suggest modifications in the models.
Incidentally, this is also an argument in favour of interpretable parametric models as opposed
to data-driven pure-regression models.
A sensible procedure to fit general purpose vision models would be alternating different learning
epochs using uncontrolled natural stimuli and well-controlled artificial stimuli to
check the biological plausibility at each point.

In conclusion, predicting subjective distances between images may
be a trivial regression problem, but using these large-scale databases to fit plausible models
may take more than that (e.g. a vision scientist in the loop doing the proper fine-tuning of interpretable models using
the classical artificial stimuli).

\section{Materials \& Methods}
\label{models}

In this section we introduce the detailed formulation of Model A and Model B and we provide links to
general purpose software that implements the models and is required for stimulus synthesis.
In addition, we introduce the specific routines that use the general purpose software to generate
the particular stimuli and the trial-and-error procedure to set the parameters leading to the responses
with the right qualitative properties.

\subsection{Generating the artificial stimuli}
\label{stimuli}

The generation of the illustrative stimuli proposed here require the generation of random noise with accurate control of
spatial frequency, luminance, contrast and appropriate rendering in the display at hand.
All this can be done using the generic routines of BasicVideoTools \citep{BasicVideoTools}.
In order to do so, one has to take into account a sensible sampling frequency
(e.g. bigger than 60 cpd to avoid aliasing at visible frequencies) and the corresponding central frequencies
and orientations of the selected wavelet filters in the model. See next subsections for detailed information
on the models and associated software.
The specific software used in this paper to generate the stimuli and to compute the response curves
is available at: {\tiny{\verb"http://isp.uv.es/docs/ArtificeReloaded.zip"}}.

\subsection{The base-line: \emph{"Model~A"}}
\label{modelA}

The starting point of our analysis is an illustrative model, here referred to as \textbf{Model~A},
that follows the program suggested in \citep{Carandini12}: a cascade of four
isomorphic L+NL layers, each focused on a different psychophysical factor:
\begin{labeling}{Layer $S^{(4)}$}
\item [Layer $S^{(1)} \,\,\,$] linear spectral integration to compute luminance and opponent tristimulus channels, and nonlinear brightness/color response.
\item [Layer $S^{(2)} \,\,\,$] definition of local contrast by using linear filters and divisive normalization.
\item [Layer $S^{(3)} \,\,\,$] linear LGN-like contrast sensitivity filter and nonlinear local energy masking in the spatial domain.
\item [Layer $S^{(4)} \,\,\,$] linear V1-like wavelet decomposition and nonlinear divisive normalization to account for
orientation and scale-dependent masking.
\end{labeling}
The mathematics of \textbf{Model~A} required to set its parameters are detailed in \citep{Martinez17}.
The software implementing this model and also the generalized \textbf{Model~B} is available at {\tiny{\verb"http://isp.uv.es/docs/BioMultiLayer_L_NL_a_and_b.zip"}}.
Note that the software for this multilayer architecture uses the CSF of the Standard Spatial Observer in \citep{Watson02}, and the
Steerable wavelet in MatlabPyrTools \citep{Simoncelli92}. Please cite these sources as well when using this code.
Below, for the reader convenience, we briefly recall the expressions of the series of forward L+NL transforms (stimulus-response).

\textbf{Model~A} represents a system, $S$, that depends on some parameters, $\vect{\Theta}$, and applies a series of transforms on the input radiance vector, $\vect{x}^0$, to get a series of intermediate response vectors, $\vect{x}^i$,
\begin{equation}
  \xymatrixcolsep{2pc}
  \xymatrix{ \vect{x}^0 \ar@/_1pc/[r]_{S^{(1)}} \ar@/^2pc/[rrrr]^{\scalebox{0.9}{$S(\vect{x}^0,\vect{\Theta})$}} & \vect{x}^1  \ar@/_1pc/[r]_{S^{(2)}}  & \vect{x}^2 \ar@/_1pc/[r]_{S^{(3)}} & \vect{x}^3 \ar@/_1pc/[r]_{S^{(4)}}  & \vect{x}^4
  }
  \label{modular}
\end{equation}
Each layer in this sequence is intended to account for the corresponding psychophysical phenomenon
outlines above. Each layer is the concatenation of a linear transform $\mathcal{L}$ and a nonlinear transform $\mathcal{N}$:
\begin{equation}
  \xymatrixcolsep{2pc}
  \xymatrix{ \cdots \vect{x}^{i-1} \ar@/_2pc/[rr]_{S^{(i)}} \ar[r]^{\,\,\,\,\,\, \mathcal{L}^{(i)}} & \vect{y}^i \ar[r]^{\mathcal{N}^{(i)}} & \vect{x}^i \cdots}
  \label{module}
  \vspace{-0.0cm}
\end{equation}
Here, in each layer we use convolutional filters for the linear part and the canonical Divisive Normalization for the nonlinear part.
The transforms performed by each specific layer are described below:

\paragraph{Layer 1: Brightness from Radiance}
\begin{eqnarray}
  \nonumber \mathcal{L}^{(1)}  \equiv \,\,\,\,\,\,\,\,\,\,\,\,\,\,\,\, \vect{y}^1 & \!\! = \!\!& L^1 \cdot \vect{x}^0 \\
  \mathcal{N}^{(1)}  \equiv \,\,\,\,\,\,\,\,\,\,\,\,\,\,\,\, \vect{x}^1 & \!\! = \!\!& K(\vect{y}^1) \cdot \mathds{D}^{-1}_{\left( \vect{b}^1 + H^1 \cdot {\vect{y}^1}^{\gamma^1} \right)} \cdot {\vect{y}^1}^{\gamma^1} \label{layer1}
\end{eqnarray}

\noindent where, $L^1$ is a matrix with the color matching functions for each spatial location. In particular, restricting ourselves to achromatic information,
the only required color matching function would be the spectral sensitivity $V_\lambda$ \cite{Stiles82,Fairchild13}, leading to the luminance in each spatial location.
The nonlinear part is the canonical Divisive Normalization, where the Hadamard products and quotients have been expressed using diagonal matrices: note
that $\vect{a} \odot \vect{b} = \mathds{D}_{\vect{a}} \cdot \vect{b} = \mathds{D}_{\vect{b}} \cdot \vect{a}$ and $\mathds{D}_{\vect{v}}$ is a diagonal matrix with the vector $\vect{v}$ in the diagonal \citep{Martinez17,Minka00}.
The global scaling matrix $K(\vect{y}^1) = \kappa \left( \mathds{D}_{\vect{b}^1} +  \mathds{D}_{\left( \frac{\beta}{d} \mathds{1} \cdot {\vect{y}^1}^{\gamma^1} \right)} + \mathbbm{1} \right)$, just ensures that the maximum brightness value (for normalized luminance equal to 1) is $\kappa$.
The role of the interaction kernel in the denominator $H^1 = \left( \frac{\beta}{d} \mathds{1} + \mathbbm{1} \right )$,
where $\mathds{1}$ is the all-ones $d\times d$ matrix, and $\mathbbm{1}$ is the identity matrix,
is setting the anchor for the brightness adaptation.
With this kernel in the denominator the anchor luminance is related to the average luminance energy $\left( \vect{b}^1 + \frac{\beta}{d} \mathds{1} \cdot {\vect{y}^1}^{\gamma^1} \right )$.
The effect of this nonlinear transform is a Weber-like adaptive saturation \cite{Abrams07}. Similar nonlinear behavior can be assumed for the opponent chromatic channels \cite{Fairchild13,Stockman11,Laparra12}, but we didnt implemented the color version of the model.

\paragraph{Layer 2: Contrast from Brightness}
\begin{eqnarray}
  \nonumber  \mathcal{L}^{(2)}  \equiv \,\,\,\,\,\,\,\,\,\,\,\,\,\,\,\, \vect{y}^2 & \!\! = \!\!& L^2 \cdot \vect{x}^1 \\
  \mathcal{N}^{(2)}  \equiv \,\,\,\,\,\,\,\,\,\,\,\,\,\,\,\, \vect{x}^2 & \!\! = \!\!& \mathds{D}^{-1}_{\left( \vect{b}^2 + H^2 \cdot \vect{y}^2 \right)} \cdot \vect{y}^2  \label{layer2}
\end{eqnarray}
\noindent where the linear stage computes the deviation of point-wise brightness with regard to the local brightness through
$L^2 = \mathbbm{1} - \mathcal{H}^n$, and this kernel in the \emph{numerator}, $\mathcal{H}^n$, represents the convolution by a two-dimensional Gaussian.
The normalization through $H^2 = \mathcal{H}^d \cdot \left( \mathbbm{1} - \mathcal{H}^n \right)^{-1}$, where the kernel in the \emph{denominator}, $\mathcal{H}^d$, is another two-dimensional Gaussian kernel, leads to the standard definition of contrast: normalization of the deviation of brightness by the local brightness.

\paragraph{Layer 3: Contrast sensitivity and spatial masking}
\begin{eqnarray}
  \nonumber \mathcal{L}^{(3)}  \equiv \,\,\,\,\,\,\,\,\,\,\,\,\,\,\,\, \vect{y}^3 & \!\! = \!\!& L^3 \cdot \vect{x}^2 \\
  \mathcal{N}^{(3)}  \equiv \,\,\,\,\,\,\,\,\,\,\,\,\,\,\,\, \vect{x}^3 & \!\! = \!\!& \mathds{D}_{\textrm{sign}(\vect{y}^3)} \cdot \mathds{D}^{-1}_{\left( \vect{b}^3 + H^3 \cdot |\vect{y}^3|^{\gamma^3} \right)} \cdot |\vect{y}^3|^{\gamma^3} \label{layer3}
\end{eqnarray}
\noindent where $L^3$ is the convolution matrix equivalent to the application of a Contrast Sensitivity Function (CSF) \cite{Campbell68}.
The rows of this matrix consist of displaced versions of center-surround (LGN-like) receptive fields (impulse response of the CSF \cite{InglingUriegas}).
The kernel in the denominator, $H^3$, represents the convolution by another two-dimensional Gaussian that computes the local contrast energy
that masks the responses in high-energy environments.

\paragraph{Layer 4: Wavelet analysis and frequency masking}
\begin{eqnarray}
  \nonumber \mathcal{L}^{(4)}  \equiv \,\,\,\,\,\,\,\,\,\,\,\,\,\,\,\, \vect{y}^4 & \!\! = \!\!& L^4 \cdot \vect{x}^3 \\
  \mathcal{N}^{(4)}  \equiv \,\,\,\,\,\,\,\,\,\,\,\,\,\,\,\, \vect{x}^4 & \!\! = \!\!& \mathds{D}_{\textrm{sign}(\vect{y}^4)} \cdot \mathds{D}^{-1}_{\left( \vect{b}^4 + H^4 \cdot |\vect{y}^4|^{\gamma^4} \right)} \cdot |\vect{y}^4|^{\gamma^4} \label{layer4}
\end{eqnarray}
\noindent where $L^4$ is the matrix of Gabor-like receptive fields corresponding to V1-like sensors \cite{SimoncelliWoods90}.
The kernel in the denominator, $H^4$, represents the masking interaction between sensors tuned to different space, frequency and orientation \cite{Watson97}.
The focus of this paper is in the effect of training this 4th stage in large-scale naturalistic databases.

Despite the reasonable formulation of this base-line \textbf{Model~A} and its successful performance in reproducing
subjective opinion in large-scale naturalistic databases, it fails in reproducing basic visual masking phenomena.

\subsection{A general and easier-to-tune solution: \emph{"Model B"}}
\label{modelB}

In this paper we focus on the modification of the last layer of \textbf{Model~A}.
As stated in the main text, \textbf{Model~B} is just \textbf{Model~A} with a number of modifications in the last nonlinear part, $\mathcal{N}^{(4)}$, in Eq. \ref{layer4},
which should account for the frequency and the orientation masking.
For the sake of clarity we will refer to the previous and modified versions of this nonlinearity as $\mathcal{N}_A$ and $\mathcal{N}_B$ respectively;
where we omit the superindices indicating the layer: in this subsection we are always referring to the 4th-layer.
As in the main text, also for simplicity, we will refer to the \emph{energy} of the linear outputs (after element-wise rectification and exponentiation) $\vect{e} = |\vect{y}|^\gamma$.

In this section we follow the two suggestions made in \citep{Martinez17} for new models:
(a) provide not only the forward transform, but also the Jacobian and the inverse; and
(b) use matrix notation.
We simply list the results for the Jacobian and the inverse with no proof, just for the reader convenience.
However, in the provided toolbox there is a routine that numerically checks the Jacobian and compares the inverse to the actual input.

\paragraph{Forward transform.}
\textbf{Model~B} consists of (a) adding a global scaling factor (diagonal matrix) to set the dynamic range of the output, and (b) generalizing the interaction kernel in the previous nonlinearity,

\begin{eqnarray}
       \mathcal{N}_B(\vect{e}) = K(\vect{e}^\star) \cdot \mathcal{N}_A(\vect{e}) & = &  K(\vect{e}^\star) \cdot \mathds{D}_{\textrm{sign}(\vect{y})} \cdot \mathds{D}^{-1}_{\left( \vect{b} + H_G \cdot \vect{e} \right)} \cdot \vect{e}
       \label{DN_B_matrix} \\
       where & &  \nonumber \\
       K(\vect{e}^\star) & = & \mathds{D}_{\vect{\kappa}} \cdot \mathds{D}_{\left( \vect{b} + H_G \cdot \vect{e}^\star \right)} \cdot \mathds{D}^{-1}_{\vect{e}^\star} \nonumber \\
       H_G & = & \mathds{D}_{\vect{c}} \cdot \left[ H_{\vect{p}} \odot H_{f} \odot H_{\phi} \odot C_{\textrm{int}} \right] \cdot \mathds{D}_{\vect{w}} \nonumber
\end{eqnarray}
where the \emph{global scaling vector}, $\vect{\kappa}$  (that determines the dynamic range of the output) is obtained from the intra-subband average of the response to natural images, $|\vect{x}|$, in \textbf{Model~A},
and the \emph{reference vector}, $\vect{e}^\star$ (that describes the dynamic range of the input to $\mathcal{N}_B$) may be independent of the input (e.g. a global normalization constant vector, also obtained from the
intra-subband average of natural images in $|\vect{y}|^\gamma$), or, similarly to Layer 1, it may depend on each specific input (auto-normalization as opposed to global normalization).
In this auto-normalization case, the dynamic range of the input may be set from the average value in each subband (as in Layer 1, where the anchor luminance depends on the average luminance).
The averages over the suubands can be computed as $\vect{e}^\star = \mathds{D}_{\frac{1}{d_w}} \cdot \mathds{1}_w \cdot \vect{e}$. Where $\mathds{1}_w$ is a block-diagonal matrix with \emph{all-ones} in the diagonal blocks corresponding to each subband and $\mathds{D}_{\frac{1}{d_w}}$ is a diagonal matrix that divides the corresponding sum by the dimension of the wavelet subband, thus leading to the average.
This apparently complicated matrix expression for the average is just to simplify the derivative with regard to the stimulus (which reduces to the constant matrix $\mathds{D}_{\frac{1}{d_w}} \cdot \mathds{1}_w$).

The generalized interaction kernel is given by the modulation (Hadamard, element-wise product) of three Gaussian kernels over the locations, $\vect{p}$, scales $f$, and orientations $\phi$ of
the wavelet-like coefficients. These Gaussian kernels have normalization constants so that they have unit volume in their definition domain.
In the case of spatial kernels, as the number of discrete samples (number of sensors per subband) depends on the subband,
the amplitude of the Gaussian kernel also depends on the subband. Given the already nonlinear nature of the input $\vect{e}$, this difference in the normalization factor
may mean that some subbands are over-weighted with regard to others. That is why we included other matrices (the full matrix $C_{\textrm{int}}$ and the diagonal matrix $\mathds{D}_{\vect{w}}$) to compensate
these effects if necessary.
Note that $\mathds{D}_{\vect{w}}$ applies column-wise weights on the final kernel, or equivalently, it selectively
weights the energy of the subbands in the input vector $\vect{e}$. This means that it can be used to moderate the effect of the
a specific subband if it is too big.
More importantly, one could act on a specific block of the full matrix, $C_{\textrm{int}}$, if the relation between two specific
subbands should be modified. We did not have to do that in this work for a qualitative fix of the behavior, i.e. $C_{\textrm{int}}$ remained an all-ones matrix.
Finally, the global normalization of each row is controlled by the vector in the diagonal matrix $\mathds{D}_{\vect{c}}$.

\paragraph{Jacobian with regard to the stimulus.}
Following \citep{Martinez17}, the Jacobian $\nabla_{\vect{x}^0} S(\vect{x}^0)$ reduces to the knowledge of the Jacobian of each stage, $\nabla_{\vect{y}^i \mathcal{N}^{(i)}(\vect{y}^i)}$,
and in our case (\textbf{Model~B}), the only term yet to be specified is, $\nabla_{\vect{y}} \mathcal{N}_B(\vect{y})$.
If we call $\vect{x}_A = \mathcal{N}_A(\vect{y},H_G) = \mathds{D}_{\textrm{sign}(\vect{y})} \cdot \mathds{D}^{-1}_{\left( \vect{b} + H_G \cdot \vect{e} \right)} \cdot \vect{e} $, we have:
\begin{eqnarray}
  \nabla_{\vect{y}} \mathcal{N} & = & K(\vect{e}^\star) \cdot \nabla_{\vect{y}} \mathcal{N}_A(\vect{y}) + \mathds{D}_{\vect{x}_A} \cdot \nabla_{\vect{y}} K(\vect{e}^\star)
  \label{jacobian_B} \\
  \nonumber \\
     where & & \nonumber \\
    \nonumber \\
  \nabla_{\vect{y}} \mathcal{N}_A(\vect{y}) & = &  \mathds{D}_{\textrm{sign}(\vect{y})} \cdot
      \mathds{D}^{-1}_{\left( \vect{b} + H_G \cdot \vect{e} \right)} \cdot \left[ \mathbbm{1} - \mathds{D}_{\left(\frac{\vect{e}}{\vect{b} + H_G \cdot \vect{e}}\right)} \cdot H_G
      \right]  \cdot \mathds{D}_{\gamma |\vect{y}|^{\gamma -1}} \cdot \mathds{D}_{\textrm{sign}(\vect{y})}   \nonumber \\
  \nabla_{\vect{y}} K(\vect{e}^\star)  & = &  \nabla_{\vect{e}} K(\vect{e}^\star) \cdot \mathds{D}_{\gamma |\vect{y}|^{\gamma -1}} \cdot \mathds{D}_{\textrm{sign}(\vect{y})} \nonumber \\
  \nabla_{\vect{e}} K(\vect{e}^\star) &=& \left \{
\begin{array}{lc}
0 & \textrm{in case} \,\, \vect{e}^\star \,\, \textrm{is constant} \\
\\
\mathds{D}_{\vect{\kappa}} \cdot \mathds{D}_{\vect{e}^\star} \cdot H_G \cdot \mathds{D}_{\frac{1}{d_w}} \cdot \mathds{1}_w + \mathds{D}_{\vect{\kappa}} \cdot \mathds{D}_{\left( \vect{b} + H_G \cdot \vect{e}^\star \right)} \cdot \mathds{D}^{-2}_{\vect{e}^\star} \cdot \mathds{D}_{\frac{1}{d_w}} \cdot \mathds{1}_w & \textrm{for auto-normalization}
\end{array}
\right.
\nonumber
\end{eqnarray}


\paragraph{Inverse.} If the reference vector, $\vect{e}^\star$, is constant, the inverse has closed form:
\begin{equation}
    \vect{y} = \mathcal{N}^{-1}_B( \, K(\vect{e}^\star)^{-1} \cdot |\vect{x}| \, ) = \mathds{D}_{\text{sign}(\vect{x})} \cdot \left[ \left( \mathbbm{1} -  \mathds{D}_{\left( K(\vect{e}^\star)^{-1} \cdot |\vect{x}| \right)} \cdot H_G  \right)^{-1} \cdot \mathds{D}_{\vect{b}} \cdot K(\vect{e}^\star)^{-1} \cdot |\vect{x}| \right]^{\frac{1}{\gamma}}
    \label{inverse}
\end{equation}
On the contrary, in the case of auto-adaptation, when coming back from certain $\vect{x}$, the reference $\vect{e}^\star$ is unknown.
Nevertheless, the inverse can still be obtained iteratively. Starting from certain guess $\vect{e}^\star_0$ (e.g. the one for natural images),
one can obtain a first guess for the inverse $\vect{y}_1$ using Eq. \ref{inverse}. From the $n$-th guess for the inverse, one can derive a new guess for the reference:
$\vect{e}^\star_n = \mathds{D}_{\frac{1}{d_w}} \cdot \mathds{1}_w \cdot |\vect{y}_n|^\gamma$, and keep the iteration:
\begin{equation}
    \vect{y}_{n+1} = \mathcal{N}^{-1}_B( K( \vect{e}^\star_{n})^{-1} \cdot \vect{x})
\end{equation}
where $\mathcal{N}^{-1}_B$ is computed using Eq. \ref{inverse}.

\vspace{-0.4cm}
\section*{Disclosure/Conflict-of-Interest Statement}

The authors declare that the research was conducted in the absence of any commercial or financial relationships that could be construed as a potential conflict of interest.

\vspace{-0.4cm}
\section*{Author Contributions}

XXX conceived the work,
prepared the data and code for the experiments, and contributed to the interpretation of the results.

\vspace{-0.4cm}
\section*{Acknowledgement}
This work was conceived in La Fabrica de Hielo (Malvarrosa) after the reaction of Dr. C.A. Parraga to \citep{Rufin17}:
scientists cannot be easily substituted by machines.
This work was partially funded by the MINECO/FEDER/EU projects CICYT TEC2013-50520-EXP
and CICYT BFU2014-59776-R, and TIN2015-71537-P;
by the European Research Council, Starting Grant ref. 306337;
and by the Icrea Academia Award.

\vspace{-0.4cm}
\section*{Supplemental Data}
The code with a Matlab implementation of both \textbf{Model~A} and \textbf{Model~B}
(that also includes numerical check of the analytical results presented in Section \ref{modelB})
is available at {\tiny{\verb"http://isp.uv.es/docs/BioMultiLayer_L_NL_a_and_b.zip"}}

\bibliographystyle{frontiersinSCNS&ENG} 
\bibliography{JesusMaloRef,references_mb,other2}

\end{document}